\documentstyle[11pt]{article}
\begin{document}
\vspace*{-2cm}
\noindent
\hspace*{11cm}
UG--FT--93/98 \\
\hspace*{11cm}
hep-ph/9808400 \\
\hspace*{11cm}
January 1999 \\
\begin{center}
\begin{large}
{\bf Constraints on top couplings in models with exotic quarks \\}
\end{large}
~\\
F. del Aguila, J. A. Aguilar--Saavedra \\
{\it Departamento de F\'{\i}sica Te\'{o}rica y del Cosmos,
Universidad de Granada \\
18071 Granada, Spain} \\
~\\
R. Miquel \\
{\it Departament ECM, Universitat de Barcelona \\
08028 Barcelona, Spain} \\
\end{center}
\begin{abstract}
The extension of the Standard Model with exotic quark singlets and doublets
introduces large flavor changing neutral couplings between ordinary fermions. We
derive inequalities which translate the precise determination of the diagonal
$Z$ couplings, in particular at LEP, into stringent bounds on the off-diagonal
ones. The resulting limits can be saturated in minimal extensions with one
vector doublet or singlet. In this case, 23 and 6 single top events,
respectively, are predicted at LEP2 for an integrated luminosity of 500
${\mathrm pb}^{-1}$ per experiment.
\end{abstract}
\hspace*{0.8cm}

\hspace*{0.8cm}

The agreement of the Standard Model (SM) with present data places strong
constraints on new physics \cite{papiro1,papiro9}. However, these are much less
stringent for the top quark. Indeed, the only model-independent limits on
anomalous top couplings are the direct ones from Tevatron,
${\mathrm Br}(t \rightarrow qZ) \leq 0.33$,
${\mathrm Br}(t \rightarrow q \gamma) \leq 0.032$ at 95\% CL \cite{papiro3}.
These limits will be improved, up to a factor $\sim$ 20 eventually, after the
Tevatron run beginning in the next millennium. One can also consider generic
indirect limits \cite{papiro2}, but they are usually too restrictive because
possible interference effects with contributions of other heavy states,
natural in many models (see Ref. \cite{papiro12b}), are not taken into account.
On the other hand, effective vertices are often suppressed by coupling
constant and mass scale factors in well-defined theories. Present Tevatron
limits are weak enough to allow for the production of $\sim 630$ clean single
top $t \bar q+\bar t q$ events at LEP2 assuming a total luminosity of 500
${\mathrm pb}^{-1}$ per experiment at 200 GeV \cite{papiro18}. However, the
expected rate of such events is negligible in the SM and many of its extensions.
Models with extra scalars (or gauge bosons) can have large flavor changing
neutral (FCN) top couplings with ordinary fermions, but their observation at
LEP2 would require a new boson with a mass close to threshold, to be exchanged
in the s-channel, and with a relatively large coupling to $e^+ e^-$. These
models, {\em e. g.} two Higgs doublet models \cite{papiro4}, need also extra
symmetries, at first sight unnatural, to allow a large and harmless boson
coupling to electrons. Then, there seems to be a gap between present direct
limits and expected single top production at $e^+ e^-$ colliders in the context
of simple SM extensions. In this Letter we point out that these events can be
produced at LEP2 in models with exotic quarks \cite{papiro12b,papiro5}, although
their rate is further constrained by a set of inequalities obeyed by these
models.

Vector-like or mirror quarks appear in many grand unified and string theories,
as those based on $E_6$. Mixing with these heavy fermions is the best way
to enhance the top signal without producing new particles if the new quarks have
masses above threshold. Heavy vector-like fermions decouple, and their indirect
effects become small under the natural assumptions of multiplet degeneracy and
mass scaling. At any rate, the induced mixing between ordinary quarks cannot be
arbitrarily large because there are stringent direct limits on the FCN couplings
of the five light flavors \cite{papiro12}. Moreover, we will derive simple
inequalities in this class of SM extensions relating the off-diagonal $Z$
couplings to the diagonal ones. They translate the precise determination of the
$u$, $c$ quark couplings into stronger constraints on $Ztq$ couplings than
present or future direct Tevatron bounds. We will prove that the limits deduced
in this way can be saturated in the simplest SM extensions with one vector-like
quark doublet or singlet, leading to a sizable production of single top events
at LEP2. For the top quark these inequalities provide a simple and novel method
to estimate to a good approximation the allowed size of the $Ztq$ coupling, with
the advantage that there is no need to perform global fits to particular models.

The class of models proposed extends the SM quark content to include vector-like
doublets (left- and right-handed doublets under $\mathrm SU(2)_L$), vector-like
singlets (left- and right-handed singlets), and mirror quarks, which are
left-handed singlets and right-handed doublets. The doublets can have electric
charges $(\frac{2}{3},-\frac{1}{3})$, $(\frac{5}{3},\frac{2}{3})$ or
$(-\frac{1}{3},-\frac{4}{3})$, although only exotic quarks with standard charges
appear in the simplest grand unified and string models, as for instance in $E_6$
with extra $27+\overline{27}$ representations. For simplicity we will first
restrict
ourselves to doublets with standard charges and generalize the results later.
Let us consider any of these extensions, with $N$ standard quark families, $n$
vector-like doublets, $n_u$ up and $n_d$ down vector-like singlets and $N'$
mirror quark families. The total number of up type quarks
${\mathcal N}_u = N+N'+n+n_u$ and down type quarks
${\mathcal N}_d = N+N'+n+n_d$ do not need to be equal in general (by `up' and
`down' we will mean charges $\frac{2}{3}$ and $-\frac{1}{3}$, and {\em not} weak
isospin $\frac{1}{2}$ and $-\frac{1}{2}$). In any of these models, the gauge
neutral current Lagrangian in the weak eigenstate basis can be written in matrix
notation as
\begin{eqnarray}
{\mathcal L}_{Z} & = & - \frac{g}{2 c_W} \left( \bar u_{L}^{(d)}
\gamma^\mu u_{L}^{(d)} + \bar u_{R}^{(d)} \gamma^\mu u_{R}^{(d)} \right.
\nonumber \\
& & \left. - \bar d_{L}^{(d)} \gamma^\mu d_{L}^{(d)} - \bar d_{R}^{(d)} 
\gamma^\mu d_{R}^{(d)} - 2 s_W^2 J_{\mathrm EM}^\mu \right) Z_\mu \,,
\label{ec:1}
\end{eqnarray}
with $(u_{L}^{(d)},d_{L}^{(d)})$ and $(u_{R}^{(d)},d_{R}^{(d)})$ doublets under
$\mathrm SU(2)_L$ of dimension $N+n$ and $N'+n$, respectively. The charged and
Higgs currents are also modified, but we ignore them for the moment. The mixing
of weak eigenstates with the same chirality and different isospin originates FCN
couplings in the mass eigenstate basis, where the Lagrangian in Eq. (\ref{ec:1})
reads
\begin{eqnarray}
{\mathcal L}_{Z} & = & - \frac{g}{2 c_W} \left(
\bar u_{L} X^{uL} \gamma^\mu u_{L} +
\bar u_{R} X^{uR} \gamma^\mu u_{R} \right. 
\nonumber \\
& & \left. - \bar d_{L} X^{dL} \gamma^\mu d_{L} - 
\bar d_{R} X^{dR} \gamma^\mu d_{R} 
 - 2 s_W^2 J_{\mathrm EM}^\mu \right) Z_\mu \,.
\label{ec:2}
\end{eqnarray}
Here $u=(u,c,t,T,\dots)$ and $d=(d,s,b,B,\dots)$ are ${\mathcal N}_u$ and
${\mathcal N}_d$ dimensional vectors, respectively. The diagonal $Zqq$ couplings
of  up and down type mass eigenstates $q=u,d$ are
$c_{L(R)}^q=\pm X_{qq}^{L(R)} -2 Q_q s_W^2$, with the plus (minus) sign for up
(down) type quarks (we will always drop $u$, $d$ superscripts if not needed). If
$q_{L(R)}$ is a pure up or down quark singlet, $X_{qq}^{L(R)}=0$, whereas for a
pure doublet $X_{qq}^{L(R)}=1$. In both cases FCN couplings between $q_{L(R)}$
and other quarks vanish for $q_{L(R)}$ is a weak interaction eigenstate and has
a definite isospin. In general $q_{L(R)}$ has singlet and doublet components and
$0 < X_{qq}^{L(R)} < 1$, implying nonzero FCN couplings for $q_{L(R)}$. These
arguments are made quantitative writing the unitary transformations between the
mass and weak interaction eigenstates, $q_{L}^0={\mathcal U}^{qL} q_{L}$,
$q_{R}^0={\mathcal U}^{qR} q_{R}$, with ${\mathcal U}^{qL}$ and
${\mathcal U}^{qR}$ ${\mathcal N}_q \times {\mathcal N}_q$ unitary matrices and
$q_{L,R}^0=(q_{L,R}^{(d)},q_{L,R}^{(s)})$ weak interaction eigenstates (doublets
$q_{L,R}^{(d)}$ and singlets $q_{L,R}^{(s)}$). Then, it follows from Eqs.
(\ref{ec:1},\ref{ec:2}) that
\begin{eqnarray}
X_{\alpha \beta}^{uL}=
({\mathcal U}^{uL}_{i\alpha})^* {\mathcal U}^{uL}_{i\beta} &,~~&
X_{\alpha \beta}^{uR}=
({\mathcal U}^{uR}_{j\alpha})^* {\mathcal U}^{uR}_{j\beta} \,, \nonumber \\
X_{\sigma \tau}^{dL}=
({\mathcal U}^{dL}_{k\sigma})^* {\mathcal U}^{dL}_{k\tau} &,~~ &
X_{\sigma \tau}^{dR}=
({\mathcal U}^{dR}_{l\sigma})^* {\mathcal U}^{dR}_{l\tau} \,,
\label{ec:4}
\end{eqnarray}
where $(i,k)$ and $(j,l)$ sum over the left- and right-handed doublets,
respectively, $\alpha,\beta=u,c,t,T,\dots$ and $\sigma,\tau=d,s,b,B,\dots$ From
these equations we obtain all the information on $Z$ couplings. To simplify the
notation, let $q,q'$ be two mass eigenstates with the same electric charge and
chirality and $X$ the corresponding coupling matrix $X^{uL}$, $X^{uR}$, $X^{dL}$
or $X^{dR}$. Eq. (\ref{ec:4}) implies that the matrix elements $X_{qq'}$ are
bounded, $|X_{qq'}| \leq 1$, and that the diagonal elements are positive, 
$X_{qq} \geq 0$. In particular, if $q$ is a weak eigenstate,
$X_{qq} = \pm 2 T_{3q}$, with the plus (minus) sign for up (down) type quarks.
Using the Schwarz inequality it is straightforward to show that for $q \neq q'$
\begin{eqnarray}
|X_{qq'}|^2 & \leq & (1-X_{qq}) (1-X_{q'q'}) \,, \label{ec:5} \\
|X_{qq'}|^2 & \leq & X_{qq} X_{q'q'} \,.
\label{ec:6}
\end{eqnarray}
These inequalities translate the determination of the diagonal $Zqq$ and $Zq'q'$
couplings, $X_{qq}$ and $X_{q'q'}$, into a bound on the off-diagonal coupling
$Zqq'$, $X_{qq'}$. In other words, Eqs. (\ref{ec:5},\ref{ec:6}) relate the
isospin of $q,q'$ with the FCN coupling $Zqq'$. Even if we do not know
$X_{q'q'}$ we still can learn about the $X_{qq'}$ coupling from our knowledge of
$X_{qq}$. This is particularly useful in the case of the top quark, since the
bounds derived can be saturated in the simplest extensions with exotic fermions.
Also we can set limits on the coupling of a light quark $q$ to a new unknown
quark $q'$. From Eqs. (\ref{ec:5},\ref{ec:6}) follows that if $X_{qq}=0,1$, the
FCN couplings involving $q$ vanish (in this case $q$ is a weak eigenstate),
independently of the particular SM extension considered. Conversely, if
$X_{qq} \neq0,1$, there must exist nonzero FCN couplings for $q$, as can be
shown from Eq. (\ref{ec:4}) observing that $X^2=X$, $X^\dagger=X$.

\begin{table}[htb]
\begin{center}
\begin{tabular}{ll}
Experimental value & $X_{qq}^{L(R)}=\pm (c_{L(R)}^q + 2 Q_q s_W^2)$ \\ \hline
$c_L^u=0.656 \pm 0.032$ & $X_{uu}^L=0.965 \pm 0.032$ \\
$c_R^u=-0.358 \pm 0.026$ & $X_{uu}^R=-0.049 \pm 0.026$ \\
$c_L^d=-0.880 \pm 0.022$ & $X_{dd}^L=1.035 \pm 0.022$ \\
$c_R^d=-0.054^{+0.154}_{-0.096}$ & $X_{dd}^R=0.209^{+0.096}_{-0.154}$  \\
\end{tabular}
\end{center}
\caption{Diagonal couplings
measured in atomic parity violation experiments. 
We use $s_W^2=0.232 \pm 0.001$
\label{tab:1} }
\end{table}

In order to apply Eqs. (\ref{ec:5},\ref{ec:6}) it is necessary to review present
experimental results on the diagonal $Z$ couplings to quarks $c_{L,R}^q$. The
$Z$ couplings to the lightest quarks $u,d$ are measured in atomic parity
violation \cite{papiro1,papiro7} and in the SLAC polarized-electron experiment
\cite{papiro8}. The determination of $c_L^u$, $c_R^u$ and $c_L^d$ is accurate,
whereas the error in $c_R^d$ is very large (see Table \ref{tab:1}). The
precision data taken at LEP and SLC provide accurate determinations of the $Zcc$
and $Zbb$ couplings at $M_Z$ (see Table \ref{tab:2}). The ratio $R_c$ is mainly
a measure of $|c_L^c|^2 + |c_R^c|^2$ and the forward-backward (FB) asymmetry
$A_{FB}^{0,c}$ of $(|c_L^c|^2 - |c_R^c|^2)/(|c_L^c|^2 + |c_R^c|^2)$. From these
data the moduli of $c_{L,R}^c$ can be extracted but not their sign. This opens
an interesting possibility that although already settled experimentally, makes
the measurement of $A_{FB}^c$ off the $Z$ pole at LEP2 very important. Of the
two sign choices for $c_R^c$, the negative value corresponds to the usual
isospin assignment for the right-handed charm quark as an almost pure singlet,
whereas the positive sign can be achieved with a large mixing ($\sim 60 \%$)
with a new right-handed doublet $T^0_R$. This ambiguity has been settled by a
combination of the low energy measurements of $A_{FB}^c$ at PEP (29 GeV) and
PETRA (35 and 44 GeV) \cite{papiro10}. The data are consistent with the negative
sign within $0.4 \sigma$, whereas the deviation is $4.2 \sigma$ for the positive
sign. Recent measurements at LEP2 \cite{papiro11} have large statistical
uncertainties but already show a preference for a negative $c_R^c$ (0.3$\sigma$)
rather than for a positive one (1.6$\sigma$). This raises the question of
whether there is any other experimental reason to exclude the large $T^0_R-c$
mixing. A large $T^0_R$ component in the $c$ quark is disfavored by the $\rho$
parameter as long as none of the $d,s,b$ quarks has a large right-handed doublet
$B^0_R$ component \cite{papiro17}. Mixing with the $d,b$ quarks would lead to
unacceptably large right-handed charged current couplings, but mixing with the
$s$ quark is allowed. (FCN currents between light quarks can be made to vanish
if only one of them mixes with the doublet \cite{papiro5}.) Only the off-peak
asymmetry for the strange quark is sensitive to the sign of $c_R^s$, but present
data $A_{FB}^{0,s}=0.131 \pm 0.035 \pm 0.013$ \cite{papiro15} do not exclude
either sign. In summary, the only strong indication that $c_R$, $s_R$ are indeed
isosinglets and do not have a large doublet component is provided by the low
energy measurement of $A_{FB}^c$. However, this results from the average of
inconclusive measurements, so a precise determination of the off-peak asymmetry
at LEP2 will be welcome. The analysis for the $b$ quark is similar: the off-peak
asymmetry and the $\rho$ parameter fix the sign ambiguity and the $B_R^0$
component in the $b$ mass eigenstate must be small \cite{papiro16}. The value of
$c_R^d$ in Table \ref{tab:1} is compatible with the two sign assignments but
again the $\rho$ parameter and the measured value of $c_R^u$ force a small
$B_R^0-d$ mixing. Table \ref{tab:3} summarizes the values of $c_{L(R)}^q$ and
$X_{qq}^{L(R)}$ obtained from $R_b$, $R_c$, $A_{FB}^{0,b}$, $A_{FB}^{0,c}$ and
their correlation matrix in Ref. \cite{papiro9} assuming small mixing with the
new quarks, as required by the SM isospin assignments.

\begin{table}[htb]
\begin{center}
\begin{tabular}{ccc}
Quantity & Data & SM Fit \\
\hline
$R_c$ & $0.1735 \pm 0.0044$ & 0.1723 \\
$A_{FB}^{0,c}$ & $0.0709 \pm 0.0044$ & 0.0736 \\
$R_b$ & $0.21656 \pm 0.00074$ & 0.2158 \\
$A_{FB}^{0,b}$ & $0.0990 \pm 0.0021$ & 0.1030
\end{tabular}
\end{center}
\caption{$R_c$, $R_b$ and asymmetries\label{tab:2} }
\end{table}

\begin{table}[htb]
\begin{center}
\begin{tabular}{ll}
Experimental value & $X_{qq}^{L(R)}=\pm (c_{L(R)}^q + 2 Q_q s_W^2)$ \\ \hline
$c_L^c=0.690 \pm 0.013$ & $X_{cc}^L=0.998 \pm 0.013$ \\
$c_R^c=-0.321 \pm 0.019$ & $X_{cc}^R=-0.013 \pm 0.019$ \\
$c_L^b=-0.840 \pm 0.005$ & $X_{bb}^L=0.996 \pm 0.005$ \\
$c_R^b=0.194 \pm 0.018$ & $X_{bb}^R=-0.039 \pm 0.018$
\end{tabular}
\end{center}
\caption{Diagonal couplings from Table \ref{tab:2}.
 For the $b$ quark we also include
the radiative correction +0.0014 to $s_W^2$
\label{tab:3} }
\end{table}

In Tables \ref{tab:1},\ref{tab:3} we observe that the values of $X_{uu}^R$,
$X_{dd}^L$, $X_{cc}^R$ and $X_{bb}^R$ are unphysical. This is worst for
$X_{bb}^R$, which is 2$\sigma$ away from the physical region $[0,1]$, a direct
consequence of the 2$\sigma$ discrepancy between the measured and the SM values
of $A_{FB}^{0,b}$. (This discrepancy can be explained in models with doublets of
charges $(-\frac{1}{3},-\frac{4}{3})$ \cite{papiro16}, where the value of
$X_{bb}^R$ is physical as discussed below.) It is then necessary a more careful
application of the inequalities, since using directly the values in Tables
\ref{tab:1},\ref{tab:3} is not appropriate. Instead, we define the $90\%$ CL
upper limit on $X_{qq'}$ as the value $x$ such that the probability of finding
$X_{qq'} \leq x$ {\em within} the physical region is 0.9. With this definition
and a Monte Carlo generator for the Gaussian distributions of $R_b$, $R_c$,
$A_{FB}^{0,b}$, $A_{FB}^{0,c}$ (correlated) and $X_{uu}^{L,R}$, $X_{dd}^{L,R}$
(assuming no correlation) we obtain the bounds in Table \ref{tab:4}, where we
also quote present direct limits \cite{papiro3,papiro12}. Alternatively, we can
shift
the unphysical values in Tables \ref{tab:1},\ref{tab:3} to the physical region
and find the $90\%$ CL upper limit as defined above, obtaining the bounds given
in parentheses.

\begin{table}[htb]
\begin{center}
\begin{tabular}{cccl}
Coupling & $X^L$ & $X^R$ & Source \\
\hline
$uc$ & $1.2 \times 10^{-3}$ & $1.2 \times 10^{-3}$ &
$\delta m_D$ \\
 & 0.033 (0.035) & 0.019 (0.028) & Inequalities \\ \hline
$ut$ & 0.84 & 0.84 & $t \rightarrow uZ$ \\
 & 0.28 (0.28) & 0.14 (0.21) & Inequalities \\ \hline
$ct$ & 0.84 & 0.84 & $t \rightarrow cZ$ \\
 & 0.14 (0.15) & 0.16 (0.18) & Inequalities \\ \hline
$ds$ & $4.1 \times 10^{-5}$ & $4.1 \times 10^{-5}$ & 
$K^+ \rightarrow \pi^+ \nu \bar \nu$ \\
 & 0.14 (0.19) & 0.62 (0.61) & Inequalities \\  \hline
$db$ & $1.1 \times 10^{-3}$ & $1.1 \times 10^{-3}$ &
$\delta m_B$ \\
 & 0.0081 (0.017) &  0.062 (0.086) & Inequalities \\ \hline
$sb$ & $1.9 \times 10^{-3}$ & $1.9 \times 10^{-3}$ &
$B^0 \rightarrow \mu^+ \mu^- X$ \\
 & 0.076 (0.11) & 0.12 (0.17) & Inequalities 

\end{tabular}
\end{center}
\caption{Experimental limits on FCN couplings and bounds deduced from
coupling inequalities in models with exotic quarks of standard charges. The
bounds in parentheses are obtained with an alternative method of estimating the
probability, explained in the text
\label{tab:4}} 
\end{table}

A few comments are in order. (i)  The more restrictive bounds on left-handed
currents are estimated using Eq. (\ref{ec:5}), whereas for right-handed currents
Eq. (\ref{ec:6}) gives a stronger constraint. (ii) These bounds are not all
independent and thus cannot be simultaneously saturated, {\em e. g.} $X_{tc}^R$
and $X_{tu}^R$ cannot have both their largest value. (iii) The couplings of the
top quark have not yet been measured, and we assume the factors $(1-X_{tt}^L)$
and $X_{tt}^R$ in Eqs. (\ref{ec:5},\ref{ec:6}) to be both equal to unity. The
factors $(1-X_{ss}^L)$ and $X_{ss}^R$ are also set equal to one because the
measurement of $A_{FB}^{0,s}$ alone does not provide any useful constraint. (iv)
The inequalities also allow to set new nontrivial limits on the FCN couplings of
the ordinary and new heavy quarks $T$, $B$. These limits can be read from Table
\ref{tab:4} making the replacement $t \rightarrow T$, $s \rightarrow B$.
It is worth to notice that the constraints on top couplings obtained from the
inequalities are more restrictive than those from top decays at Tevatron.

The limits obtained on top FCN couplings can be saturated in the simplest
extensions of the SM, namely the addition of a vector-like doublet or singlet.
In the model with an additional isodoublet the bound $X_{ct}^R \leq 0.16$ can be
saturated choosing the projection of the new right-handed doublet $T^0$ on the
mass eigenstates $u,c,t,T$ to be
${\mathcal U}_{T^0 \alpha}^{uR}=(0,0.16,0.99,0)$. The projection of its partner
$B^0$ on the mass eigenstates can be chosen as
${\mathcal U}_{B^0 \sigma}^{dR}=(0,\epsilon,0,\sqrt{1-\epsilon^2})$. The $\rho$
parameter prefers a sizeable $B^0-s$ mixing $\epsilon$ but $b \rightarrow s
\gamma$ requires a negligible value $O(10^{-3})$. Then the
experimental constraints on FCN currents are satisfied and the right-handed
currents between ordinary quarks remain small. The analysis for
$X_{ut}^R \leq 0.14$ can be performed similarly. These bounds lead to 23 and 18
$t\bar q+ \bar t q$ events, respectively, at LEP2.
 
In the extension of the SM with one extra singlet $T^0$ of charge $\frac{2}{3}$,
there are no right-handed charged currents and all FCN couplings vanish except
those for left-handed up type quarks. This model and the analogous model with an
extra charge $-\frac{1}{3}$ singlet have been analyzed extensively in the
literature \cite{papiro12b,papiro13}, and we will not repeat their discussion
here. However, it is worth to note that in these models the relationship
between charged and neutral currents further restricts the allowed size of the
FCN couplings. Present limits on CKM matrix elements \cite{papiro1} imply
$X_{ct}^L \leq 0.082$, $X_{ut}^L \leq 0.046$, leading to 6 and 2
$t\bar q+\bar t q$ events at LEP2.

Now we will extend the analysis with the inclusion of quark doublets with
nonstandard charges $(\frac{5}{3},\frac{2}{3})$ or
$(-\frac{1}{3},-\frac{4}{3})$, which we will simply refer to as `nonstandard
doublets'. The weak interaction eigenstates are in this case
$q_{L,R}^0=(q_{L,R}^{(d)},q_{L,R}^{(n)},q_{L,R}^{(s)})$, with $q_{L,R}^{(n)}$
the nonstandard doublets which have diagonal couplings
$-\bar u_{L,R}^{(n)} \gamma^\mu u_{L,R}^{(n)}$ and
$+\bar d_{L,R}^{(n)} \gamma^\mu d_{L,R}^{(n)}$ in Eq. (\ref{ec:1}).
The form of the Lagrangian in the mass eigenstate basis remains the same
for the quarks with standard charges, counting the new quarks in
${\mathcal N}_u$ and ${\mathcal N}_d$. One major difference is that this time
the diagonal couplings are not positive definite, but still $|X_{qq'}| \leq 1$,
and $X_{qq} = \pm 2 T_{3q}$ for states with definite isospin. Moreover, while
Eq. (\ref{ec:5}) remains true, Eq. (\ref{ec:6}) is no longer valid and must be
replaced by
\begin{eqnarray}
|X_{qq'}|^2 & \leq & (1+X_{qq}) (1+X_{q'q'}) \,.
\label{ec:9}
\end{eqnarray}
The reason of this replacement becomes clear if we consider the meaning of Eqs.
(\ref{ec:5},\ref{ec:6}). For instance, applied to up type quarks they express
the fact that a mass eigenstate with the highest ($\frac{1}{2}$) or the lowest
(0) isospin is a weak eigenstate whose FCN couplings must vanish. With the
addition of nonstandard doublets, the lowest isospin for charge $\frac{2}{3}$
quarks is $-\frac{1}{2}$, thus the replacement of Eq. (\ref{ec:6}) by Eq.
(\ref{ec:9}). (For charge $-\frac{1}{3}$ quarks the argument is similar.) The
bounds obtained from Eq. (\ref{ec:9}) are much less restrictive and clearly
cannot be saturated due to the other experimental constraints (charged currents,
neutral meson mixing and oblique corrections), and a global fit is needed to
find
the largest allowed value of the FCN couplings in each particular model.
However, the bounds on $X_{qq'}^L$ couplings in Table \ref{tab:4} obtained using
Eq. (\ref{ec:5}) are still valid, as long as the highest (lowest)
isospin for left-handed up (down) quarks is $\frac{1}{2}$ ($-\frac{1}{2}$). 

Finally, some concluding remarks. If the top quark is indeed observed at LEP2,
this will provide a strong indication for the existence of new heavy quark
doublets, because the maximum $t \bar q + \bar t q$ production rate with only
new heavy singlets is four times smaller. In any case, the Next Linear Collider
with its expected integrated luminosity of 100 ${\mathrm fb}^{-1}$ at
$\sqrt s=500$ ${\mathrm GeV}$ will allow to disentangle $X_{tq}^L$ and
$X_{tq}^R$ \cite{papiro20}. On the other hand, the absence of top events at LEP2
will improve the limits on $X_{tq}^{L,R}$ to a common value of $0.033$. This
number is to be compared with the bound obtained after the next Tevatron run if
the decay $t \rightarrow qZ$ is not observed, eventually $|X_{tq}^{L,R}| \leq
0.19$. Thus, in the next two years LEP2 will either detect $Ztc$ couplings or
set on them the most stringent bounds before the next generation of colliders.
In fact the first quoted results at LEP2 \cite{papiro19} improve by a factor of
2 the present Tevatron limit.

\vspace{1cm}
\noindent
{\Large \bf Acknowledgements}

\vspace{0.4cm}
\noindent
This work was partially supported by CICYT under contract AEN96--1672 and by the
Junta de Andaluc\'{\i}a.

\end{document}